\documentclass[amstex]{article}

\usepackage{latexsym}
\usepackage{amsmath,amscd}
\usepackage{amssymb}
\usepackage{amsfonts}
\usepackage{graphicx,psfrag}
\usepackage{epsfig}

\makeatletter  

\@addtoreset{equation}{section}

\makeatother

\newcommand{\calA}{{\cal A}}
\newcommand{\calB}{{\cal B}}

\newcommand{\calL}{{\cal L}}

\newcommand{\calO}{{\cal O}}

\newcommand{\tr}{{\rm tr}}

\newtheorem{theorem}{Theorem}[section]
\newtheorem{proposition}[theorem]{Proposition}
\newtheorem{lemma}[theorem]{Lemma}
\newtheorem{corollary}[theorem]{Corollary}
\newtheorem{remark}[theorem]{Remark}

\newtheorem{definition}[theorem]{Definition}

\newcommand{\proof}{\noindent {\em Proof:~}}  

\newcommand{\nn}{\nonumber} 
\newcommand{\bR}{{\mathbb R}}
\newcommand{\bN}{{\mathbb N}}

\newcommand{\bZ}{{\mathbb Z}}

\def\bd{\begin{displaymath}}
\def\ed{\end{displaymath}}

\def\eqref#1{(\ref{#1})} 
\def\qed{\hbox{\hskip 6pt\vrule width6pt height7pt depth1pt
    \hskip1pt}\bigskip}  

\def\to{\rightarrow}

\def\runinend{\enspace}
\def\ackname{Acknowledgment\runinend}%
\def\acknowledgments{\par\addvspace{17pt}\rmfamily
\def\ackname{Acknowledgments\runinend}%
\trivlist\if!\ackname!\item[]\else
\item[\hskip\labelsep
{\bf\ackname}]\fi}%
\newcommand{\all}[1]{{\mathfrak A}_{(-\infty, -1]}}

\begin{document} 
\bibliographystyle{plain} 

 
\thispagestyle{empty}

\begin{center}{\bf \Large Ruelle-Lanford functions for quantum spin systems}
\vspace{6mm}

\large{\bf Yoshiko Ogata\footnote{email: yoshiko@ms.u-tokyo.ac.jp. Supported by JSPS 
Grant-in-Aid for Young Scientists (B), Hayashi Memorial Foundation for Female Natural 
Scientists, Sumitomo Foundation, and Inoue Foundation.}}

{\small\it Graduate School of Mathematics, University of Tokyo, Japan \\ }

\vspace{5mm}

\large{\bf Luc Rey-Bellet\footnote{email: luc@math.umass.edu, Supported by NSF, Grant DMS-0605058.}}

{\small\it Department of Mathematics and Statistics, University of 
Massachusetts, USA \\ }

\end{center}

\vspace{-1mm}

\noindent
\begin{abstract} We prove a large deviation principle for the expectation of macroscopic
observables in quantum (and classical) Gibbs states.   Our proof is based on Ruelle-Lanford
functions \cite{La,Ru65} and direct subadditivity arguments, as in the classical case \cite{LPS2,Pf},
instead of relying on G\"artner-Ellis theorem, and cluster expansion or transfer operators as done 
in the quantum case in \cite{LLS,GLM,NR,LRB,HMO,Og}.  In this approach we recover,  
expand, and unify quantum (and classical) large deviation results for lattice Gibbs states.  In the 
companion paper  \cite{OR} we discuss the characterization of rate functions in terms of 
relative entropies. 
\end{abstract}

\noindent 
\bigskip
\hrule
\bigskip

\section{Introduction}

In a large physical system in thermal equilibrium,  macroscopic observables,  such as the energy per 
unity volume,  magnetization per unit volume, and so on, have, as a rule, a distribution which is very 
sharply concentrated around their equilibrium mean value. The fluctuations of such observables 
are expected to be exponentially small in the volume $|\Lambda|$ of the physical domain except at a 
first phase order phase transition where coexisting phases can induce macroscopically large fluctuations. 

In classical mechanics systems this problem is mathematically very well-understood and very general 
{\em large deviations theorems} have been proved both for systems on a lattice or in the continuum
see \cite{La,Ru65,Ol,FO,Co,DSZ,Ge1,Ge2,LPS2,Pf,RZ}. 

For quantum mechanical systems the problem of large deviations has, in comparison, received 
little attention and is only partially understood.   The difficulty lies, partly,  in  the non-commutativity 
of quantum  mechanical observables but at a deeper level, in the   lack of control on the boundary  
effects in  quantum mechanics.   Known bulk/boundary estimates  are sufficient to prove the existence 
of thermodynamic functions see e.g., \cite{Ru,Is,BR,Si}  but they are, so far,  {\em not} sufficient to prove 
general large deviation results, especially at low temperatures for spatial dimension more than $1$.  
A number of quantum large deviation  results have been  proved in the past few 
years \cite{PRV,LLS,GLM,LRB,NR,HMO,HMOP,DMN,Og,DMNR}, (see also \cite{BDKSSR2} for an 
information-theoretic interpretation of relative entropy). Common to all these  papers is that the 
large deviation results are obtained by an application of  G\"artner-Ellis theorem.  In this paper we take 
an alternative route to large deviation results using  direct  subadditivity arguments which go back to 
the seminal paper of Lanford \cite{La} based on previous results by Ruelle \cite{Ru65} (for a clear 
exposition of this approach in our context and further references  see \cite{LPS2}).  This approach 
is particularly useful if the logarithmic moment generating functions  (i.e., a suitable free energy 
functional) lack  smoothness in which case G\"artner-Ellis theorem cannot  be applied directly. 
The subadditivity  argument, by comparison, provide automatically the large 
deviation lower bounds.  In this paper, using this approach,  we recover, unify, and extend the known 
large deviation results for quantum  (and classical) spin systems.  In addition the proofs given here 
are short and self-contained.

This paper is organized as follows. In section 2 we give a brief exposition of the road to large deviation
via subadditivity arguments, which amounts to proving the existence of the Ruelle-Lanford function which 
is an Boltzmann entropy-like functional.  In section 3 we recall the elements of the quantum spin 
system formalism needed in the paper and we introduce the bulk/boundary estimates needed on the state 
of quantum systems.  In section 4 we prove large deviation theorems for four different cases:  
(a) Commuting observables, (b) Classical observables, (c) Finite-range observables in dimension $1$. 
The discussion of the rate functions and 
their characterization in terms of relative entropies is in the  companion paper \cite{OR}.

\section{Ruelle-Lanford functions}

Let $X$ be a complete metric space, let $\{\mu_n\}$ be a sequence of Borel probability 
measures on $X$, and let  $\{v_n\}$ an increasing sequence of positive numbers 
with $\lim_{n\to \infty} v_n = +\infty$.    We say that $\mu_n$ satisfies a {\em large deviation
principle (LDP)} on the scale $v_n$ if there exists a function $I : X \to [0, \infty]$,  lower
semicontinuous and with  compact level sets, such that  for any closed set $C$ 
\begin{equation}\label{upb}
{\displaystyle \limsup_{n\to \infty} \frac{1}{v_n} \log\mu_n(C) \le -\inf_{x\in C} I(x)} \,,
\end{equation}
and for any open set $O$
\begin{equation}\label{lwb}
\displaystyle  -\inf_{x \in O} I(x) \le \liminf_{n\to \infty} \frac{1}{v_n}
\log\mu_n(O) \,.
\end{equation}
The function $I$ is called the {\em rate function} for the LDP.

In statistical mechanics applications the measures $\mu_n$ are often distributions of 
sums of  $\bR-$ or $\bR^d$- valued  weakly  dependent random variables.  
One standard approach to prove an LDP is to combine the exponential Markov 
inequality for the upper bound  \eqref{upb} and a change of measure 
and ergodicity argument for the lower bound \eqref{lwb} (see e.g. the proofs of Cramer
and G\"artner-Ellis theorem in \cite{DZ}).  
In the presence of phase transitions, i.e., lack of ergodicity with respect to spatial translation,  
additional arguments are  needed to provide a lower bound.  For example, in \cite{FO}, the 
lower bound for the LDP  for classical lattice Gibbs states is obtained by using the Shannon 
McMillan theorem  and an approximation argument by ergodic states.

Another route to LDP's using subadditivity arguments, much in the spirit 
of statistical mechanics,  was pioneered in a remarkable paper by Lanford \cite{La}, itself 
based on  earlier work by Ruelle \cite{Ru}.  We follow closely here the presentation in  \cite{LPS2}, 
see also  \cite{Pf}. 

For Borel sets $B$ let us define the set functions      
\begin{eqnarray}
\overline{m}(B) \,= \, \limsup_{n\to \infty} \frac{1}{v_n} \log \mu_n(B) \,, \quad \underline{m}(B) \,=\, 
\liminf_{n\to \infty} \frac{1}{v_n} \log \mu_n(B) \,.
\end{eqnarray}
One has the elementary properties
\begin{enumerate}
\item For any Borel set $B$ we have  $-\infty \le \underline{m}(B) \le \overline{m}(B) \le 0$. 
\item If $B_1 \subset B_2$ then $\underline{m}(B_1) \le \underline{m}(B_2)$ and 
$\overline{m}(B_1) \le \overline{m}(B_2)$. 
\item For all $B_1$, $B_2$ we have $\overline{m}(B_1 \cup B_2) \,=\, \max \{ 
\overline{m}(B_1), \overline{m}(B_2) \}$. 
\end{enumerate}
The property 3 is an key property in large deviations and is usually refereed to as the 
{\em principle of the largest term}: large deviations occur in the least unlikely way of all possible ways.

Let $B_\varepsilon(x)$ denote the ball of radius $\varepsilon$ centered at $x$
and let us define  
\begin{equation}
\overline{s}(x)\,=\, \inf_{\varepsilon}  \overline{m}(B_\varepsilon(x)) \,,  \quad
\underline{s}(x) \,=\, \inf_{\varepsilon}  \underline{m}(B_\varepsilon(x)) \,.
\end{equation}

\begin{definition}
The pair $(\mu_n, v_n)$ has a {\em Ruelle-Lanford function} (RL-function)
$s(x)$ if 
$$
\overline{s}(x) \,=\, \underline{s}(x) \,,
$$
for all $x \in X$. In this case we set $s(x) \,=\, \overline{s}(x) \,=\, \underline{s}(x)\,.$
\end{definition}
The next proposition is standard and shows that the existence of RL-function 
(almost) implies the existence of a  LDP.

\begin{proposition}\label{wldp} The Ruelle-Lanford function $s(x)$ is upper semicontinuous
and
\begin{eqnarray}
\underline{m}(O) \,&\ge& \, \sup_{x \in O} s(x)\,,  \quad O {\rm ~open}  \, \\
\overline{m}(K) \,&\le&\, \sup_{x \in K} s(x) \,, \quad K {\rm ~compact}
\end{eqnarray}
\end{proposition}

\proof ({\it sketch}) The upper semicontinuity  follows from the definition. The lower bound is 
immediate: For any  $x\in O$ and $\varepsilon$ 
sufficiently small we have 
$\underline{m}(O) \ge \underline{m}(B_\varepsilon(x))$ and thus $\underline{m}(O) 
\ge \underline{s}(x)=s(x)$ for all $x\in O$.   

To prove the upper bound, given $\varepsilon >0$ we cover the compact set $K$ by $N=N(\varepsilon)$  
balls $B_\varepsilon(x_l)$ with centers in $x_l \in K$. Using properties 2. and 3. we have 
$$
\overline{m}(K) \le  \overline{m}(\cup_{l=1}^N B_\varepsilon(x_l)) \le \max_{l} \overline{m}( 
B_\varepsilon(x_l)) \le \sup_{x \in K} \overline{m}(B_\varepsilon(x))\,.
$$
Since $\varepsilon$ is arbitrary the upper bound follows.  \qed

The statement in Proposition \ref{wldp} is usually referred to as a {\em weak large deviation
 principle} since the upper bound holds only for compact sets.  In the problems discussed in this
 paper  the probability measures $\mu_n$ are supported  uniformly on compact sets and the
 previous lemma yields immediately a large deviation principle with rate function $-s(x)$. More 
 generally one obtains a  large deviation principle by combining Proposition \ref{wldp} with a 
proof that the sequence of probability measures ${\mu_n}$ is {\em exponentially tight} (see 
e.g. \cite{DZ}, Section 1.2).  

To identify the rate function we use a standard large 
deviation result

\begin{proposition} (Laplace-Varadhan's Lemma).  Suppose that $\mu_n$ satisfies a large 
deviation principle on the scale $v_n$ with rate function $I(x)$.  Let $f$ be any continuous 
function and suppose that for some $\gamma >1$ we have the moment condition 
$\limsup_{n \to \infty} \frac{1}{v_n} \log  \mu_n\left(   e^{\gamma v_n f(x)} \right)   \,< \, \infty $. Then  
$$
\lim_{n \to \infty} \frac{1}{v_n} \log \mu_n\left(  e^{v_n f(x)} \right) \,=\, \sup_{x} \left(  f(x) -I(x) \right) \,.
$$
\end{proposition}

If $X=\bR^n$ and $f(x)= \alpha \cdot x$ we obtain
$$
e(\alpha)\,\equiv\,\lim_{n \to \infty} \frac{1}{v_n} \log \mu_n\left(  e^{v_n \alpha \cdot x} \right) \,=\, 
\sup_{x} \left( \alpha \cdot x + s(x) \right) \,,
$$
i.e., the moment generating function of $\mu_n$ is the Legendre transform of $-s(x)$.  
If, in addition, {\em we know, a priori,} that the rate function $s(x)$ is {\em concave} 
then by convex duality we obtain that 
$$
s(x) \,=\, \inf_{\alpha} \left( e(\alpha) - \alpha \cdot x \right) \,,
$$
that is, the rate function is the Legendre transform of the logarithmic moment generating function. 
Note that in our examples the moment condition will be trivially satisfied. 

\section{Quantum lattice systems}

\subsection{Interactions and states}\label{formalism}  
We introduce some notations and briefly recall  the mathematical framework for quantum 
spin systems, \cite{Is,Si,BR,AM}. 

\medskip

\noindent {\bf $C^*$-algebras:} Let $\cal A$ be a finite-dimensional $C^*$-algebra. For any 
finite  subset 
$\Lambda \subset \bZ^d$ let $\calO_\Lambda= \otimes_{x \in \Lambda} \calO_x$ 
where $\calO_x$ is isomorphic to $\calA$.  If $\Lambda \subset \Lambda'$, there is 
a natural embedding $\calO_\Lambda$ into $\calO_{\Lambda'}$  and the algebras 
$\{ \calO_\Lambda \}_{\Lambda \subset \bZ^d\,,\, {\rm finite}}$  form a partially ordered 
family of matrix algebras.  The {\em algebra of observables} for the infinite
system is given by the $C^*$-inductive limit ${\cal O}$ of  $\cup_{\Lambda \subset \bZ^d\,,\, 
{\rm finite}} \calO_\Lambda$. 

\medskip

\noindent{\bf States:} 
Let $\omega$ be a state on $\calO$, i.e., $\omega$  is a positive, normalized linear functional on
$\calO$.  Let $\{\tau_x\}_{x \in \bZ^d}$ denote the group of spatial translations.  A state $\omega$ is 
called {\em translation invariant} if $\omega (\tau_x A
) = \omega(A)$ for all $x \in \bZ^{d}$ and all $A \in \calO$. The
action of $\bZ^d$ on $\calO$ is asymptotically abelian \cite{BR} and thus the
set of translation invariant states is a simplex.  We say that a state is {\em ergodic} if it is an 
extremal point of this simplex.

\medskip

\noindent{\bf Classical subalgebras and states:}  A standard probabilistic setting is recovered 
by considering commutative (sub)algebras.  Let ${\cal A}^{(cl)}$ be an abelian subalgebra of 
${\cal A}$ with $N= \dim {\cal A}^{(cl)}$.  
For finite subsets $\Lambda$ of $\bZ^d$ let $\calO^{(cl)}_\Lambda= 
\otimes_{x \in \Lambda} \calO^{(cl)}_x$ with $\calO^{(cl)}_x$ is isomorphic to ${\calA}^{(cl)}$. 
We denote  by ${\cal O}^{(cl)}$ the inductive limit of  
$\cup_{\Lambda \subset \bZ^d\,,\,  {\rm finite}} \calO^{(cl)}_\Lambda$.  
The commutative algebra   $\calO^{(cl)}$ can be identified with $C({\cal L})$ where 
${\cal L} =\{ 1, \cdots, N\}^{\bZ^d}$ 
with product toporogy is called a {\em classical} $C^*$-algebra.  The restriction 
of any state $\omega$ on ${\cal O}$  gives a normalized linear functional $\omega^{(cl)}$ on 
$\calO^{(cl)}$.  By Riesz Markov  Theorem there exists  a probability measure 
$d\omega^{(cl)}$ such that for any $A\in \calO^{(cl)}$ 
$$
\omega(A) \,=\, \omega^{(cl)}(A) \,=\, \int_{\cal L} A( l ) d\omega ({l}) \,.
$$

\medskip
\noindent{\bf Interactions and Hamiltonians:} An {\em interaction} 
$\Psi = \{ \psi_X \}_{X \subset \bZ^d \,,\, {\rm finite}}$ is a map from the the finite subsets 
of  $\bZ^d$ to selfadjoint elements $\psi_X$ in $\calO_X$.  We will assume throughout this 
paper that $\Psi$ is translation invariant, i.e., $\tau_x( \psi_X)= \psi_{X+x}$ for any 
$X \subset \bZ^d$ and any $x \in \bZ^d$.  An  interaction $\Psi$ is {\em classical} if 
there exists a classical  $C^*$-subalgebra $\calO^{(cl)}$ such that 
$\psi_X \in \calO^{(cl)}$ for all $X \subset \bZ^d$. 

We equip translation invariant interactions $\Psi$ with the norm 
$$\|\Psi\| \equiv \sum_{X \ni 0} |X|^{-1} \|\psi_X\| \,, 
$$
where $|X|$ is the cardinality of the set $X$  and denote by $\calB$ the corresponding 
Banach space. To any interaction $\Psi\in \calB$ we associate {\em Hamiltonians}  (or 
{\em macroscopic observables})  $K_\Lambda=K_\Lambda(\Psi)$:  For  $\Lambda \subset \bZ^d$ 
finite we define 
$$
K_\Lambda = \sum_{X \subset \Lambda} \psi_X\,.
$$
Furthermore to any $\Psi \in \calB$ we associate an observable in $\calO$ by
$$
A_\Psi \,=\, \sum_{X \ni 0} \frac{1}{|X|} \psi_X \,.
$$
When we consider Gibbs state, two kinds of interactions
$\Psi$ and $\Phi$ will be considered.The interaction $\Psi$
corresponds to the observables while $\Phi$ defines the Gibbs state.
We denote by $K_\Lambda$ the local Hamiltonian associated with $\Psi$
and by $H_\Lambda$ associated with $\Phi$.

\medskip
\noindent{\bf Large deviations:} For $n\in \bN$ let 
$\Lambda(n) = \{ z \in \bZ^d \,;\, 0 \le z_i \le n-1 \}$  denote the cube  with $|\Lambda(n)|=n^d$ 
lattice points and left hand corner at the origin.  If $\omega$ is an ergodic state then the von 
Neumann ergodic theorem implies that 
$$
\lim_{n \to \infty} \frac{1}{|\Lambda(n)|} K_{\Lambda(n)} \,=\,  \omega( A_\Psi)
$$
strongly in the GNS  representation and it is natural to investigate the large  deviation properties, 
on the scale $v_n = |\Lambda(n)|$, of the sequence of Borel measures on $\bR$ 
\begin{equation*} \label{bpm}
\mu_n(A)\,\equiv\, \omega\left( {\bf I}_{A} \left(
|\Lambda(n)|^{-1} K_{\Lambda(n)}\right)\right)
\end{equation*}
where $A$ is a Borel set and  ${\bf I}_A(H)$ denotes the
spectral projection onto the eigenspace of $H$ spanned by the eigenvalues
contained in the set $A$.  We interpret the  $\mu_n(A)$ as  the  probability that 
the observables  $ |\Lambda(n)|^{-1} K_{\Lambda(n)}$ takes value in $A$ if the system 
is in the state $\omega$.

\subsection{Asymptotically decoupled states}

The states we consider in this paper obey a property of  weak dependence between disjoint 
regions of the lattice. We follow here  the terminology used in \cite{Pf} for the classical case.  

Let $C(m)$ be an arbitrary cube of side length $m$ and let us denote by $C^r(m)$ 
the cube of side length $m+2r$ centered at the same point of $\bZ^d$ as $C(m)$. 

\begin{definition} 
A state $\omega$ on $\calO$ is {\em asymptotically decoupled} with
parameters $g$ and $c$ if
\begin{enumerate}
\item  There exist a function $g: \bN \to \bN$ with $\lim_{m \to \infty} 
g(m)/m =0$  and a function 
$c \,:\, \bN \to [0,\infty)$ with $\lim_{m \to \infty} c(m)/ |C(m)| =0$. 
\item For any cube $C(m)$, $m \in \bN$, any nonegative $A \in \calO_{C(m)}$, 
any nonnegative $B \in \calO_{ C^{g(m)}(m)^c }$ we have 
$$
e^{-c(m)} \omega(A) \omega(B) \,\le\, \omega(AB)  \,\le\, e^{c(m)} \omega(A) 
\omega(B) \,.
$$
\end{enumerate}
\end{definition}
Examples of asymptotically decoupled states are 

\vspace{2mm} 
\noindent
{\bf (a) Product states.} Any product state $\omega_0$ is asymptotically decoupled 
with parameters $c=g=0$. 

\vspace{2mm} 
\noindent
{\bf (b) Classical Gibbs states.}  Let $\calO^{(cl)}$ be a classical
$C^*$-algebra and let $\Phi$ be a classical translation invariant
interaction such that $\|\Phi\|_0 \equiv \sum_{X\ni 0} \|\phi_x\|$ is
finite.  A Gibbs state for the interaction $\Phi$ is a probability measure
$\omega^{(\Phi)}$ which satisfies the DLR equation (see e.g. \cite{Ru,Si}). 
Using the DLR equation one proves easily (see e.g. \cite{LPS2}, Section 9)  
that for any positive $A \in {\cal O}_{C(m)}$ we have
\begin{equation}\label{decoup0cl}
e^{-c(m)} \omega^{(\Phi)} (A)  \,\le \, \frac{ \tr(A e^{- H_\Lambda}) }{\tr(e^{-H_\Lambda})} \,\le\, e^{c(m)}
\omega^{(\Phi)} (A) 
\end{equation}
with $c(m) = \| W_{C(m)} \|$  where $W_{C(m)}$ is the boundary interaction
$ W_{C(m)} \,=\, \sum_{X \cap C(m) \not= \emptyset \atop 
X \cap C(m)^c\not= \emptyset} \phi_X $. This implies easily that $\omega^{(\Phi)}$ is 
asymptotically decoupled if $\|\Phi\|_0 < \infty$. 

\vspace{2mm} 
\noindent
{\bf (c) Quantum KMS states.}  Let $\Phi$ be a translation invariant interaction.  A 
KMS state for the interaction $\Phi$  is a state which satisfies the KMS condition or 
equivalently the Gibbs condition  which is a quantum analog of the DLR equation
(see e.g. \cite{BR,Si} and \cite{AM} for an up-to-date presentation)
It is {\em not known} if KMS-Gibbs states are asymptotically decoupled, in general.  
Let us assume however that  \cite{Ar1,Ar2} either

\vspace{2mm}
\noindent
(i) $d=1$ and $\Phi$ finite range (i.e.,  for some $R>0$ ${\rm diam X}>R$ implies $\phi_X=0$), 

\vspace{1mm}
or 
\vspace{1mm}

\noindent
(ii) $d$ arbitrary and $\|\Phi\|_\lambda \equiv \sum_{X \ni 0} e^{\lambda |X|} \|\phi_X\|$ 
is sufficiently small, \\
then one can show that for a Gibbs-KMS state $\omega^{(\Phi)}$  and $A \in {\cal O}_{C(m)}$ we have 
the bound \eqref{decoup0cl}
where $c(m) = C(\Phi) \sum_{X \cap C(m) \not= \emptyset \atop  X \cap C(m)^c\not= \emptyset} \|\phi_X\|$. 
Contrary to the classical case the bound is highly nontrivial to prove and relies on the Gibbs condition,
Araki perturbation theory, and control of imaginary-time dynamics. This bound implies that 
$\omega^{(\Phi)}$ is asymptotically decoupled.

\vspace{2mm}
\noindent
{\bf (d)Markov measures.} Let $\omega$ be a stationary Markov chain on
a finite state space with transition matrix $Q$ and invariant
probability $q$. Then $\omega$ is asymptotically decoupled if and only if $Q$ is
irreducible and aperiodic (i.e. mixing).  If $m$ is the smallest integer such that
$Q^m$ has strictly positive entries then the parameters are
$$
g(m)=m-1\,, \quad c(n) = \sup_{\sigma_1, \sigma_2} \left| 
\log \frac{ Q^m ( \sigma_1, \sigma_2) }{ q( \sigma_2)} \right| \,. 
$$

\vspace{2mm} 
\noindent
{\bf (e) Finitely correlated states.}  These states are a non-commutative generalization of 
Markov measures and are asymptotically decoupled if and only if they are mixing which 
occur under suitable conditions similar to the aperiodicity condition 
for Markov measures.  See \cite{HP, FNW,Og} for details.

\section{Quantum large deviations theorems} 

We prove several large deviations theorems for quantum states 
(in order of increasing difficulty) by showing the  existence of concave RL-functions.  
This unifies, simplifies and  extend  a number of quantum large deviation  results  which have 
been proved with different techniques  (G\"artner-Ellis Theorem via transfer operators, 
cluster expansions, etc..).  Our proof have the advantage of being fairly short, 
self-contained,  to apply in some situations where the  rate function is not smooth.

\subsection{Preliminaries}\label{preliminaries}

In this section we prove an energy estimate used throughout the paper and explain the 
strategy (after \cite{LPS2}) used to  prove the existence of a concave Ruelle-Lanford function. 

The first fact is a very slight variation on standard bulk/boundary energy estimate, 
see e.g. \cite{Si,BR,Pf}.   Given integers $n$ and  $m$   and a function $g(m)$ such that 
$\lim_{m\to \infty} g(m)/m =0$ we  choose $k$ to be largest {\em even} integer such that 
\begin{equation*}
n = k ( m + 2g(m)) + r  \,, \quad \quad 0 \le r < 2 (m+2g(m))  \,,
\end{equation*}
(having $k$ even will be convenient in the sequel).    
We next decompose the cube $\Lambda(k(m+2g(m))$ into $k^d$ pairwise disjoint and 
contiguous cubes ${\tilde C}_j$, each of which are each translates of $\Lambda(m + 2g(m))$ 
and then further divide each cube  $\tilde{C}_j$ into a cube $C_j$ which is centered at the 
same point as $\tilde{C}_j$  and is a translate of $\Lambda(m)$ and a "corridor" 
$\tilde{C}_j \setminus C_j$ of width $g(m)$.    We shall need estimates on the difference 
between  the Hamiltonian $K_\Lambda(n)$ and the "decoupled" Hamiltonian 
for the collection of cubes $C_j$,  i.e, $\sum_{j=1}^{k^d}K_{C_j}$.

\begin{lemma} \label{simple} Let $\Psi$ be an interaction with  
$\|\Psi\|\equiv \sum_{X\ni 0} |X|^{-1} \|\psi_x\| < \infty$.  
Then there exists a function  $F(m)= F(m,\Psi)$ with 
$
\lim_{m \to \infty} F(m) \,=\, 0 
$, 
such that 
\begin{equation}
\limsup_{n\to \infty} \frac{1}{|\Lambda(n)|}\left\| K_{\Lambda(n)} - \sum_{j=1}^{k^d}K_{C_j} \right\|  \,\le\,  F(m) \,.
\end{equation}
\end{lemma}

We will also use an immediate consequence of Lemma \ref{simple}
 
\begin{corollary}\label{verysimple} Let $\Psi$ be an interaction with  
$\|\Psi\| < \infty$.   Then there exists a function  $F(m)= F(m,\Psi)$ with 
$
\lim_{m \to \infty} F(m) \,=\, 0 
$, such that
\begin{equation}
\limsup_{n \to \infty} \left\| \frac{1}{|\Lambda(n)|}K_{\Lambda(n)} - 
\frac{1}{|\Lambda(km)|} \sum_{j=1}^{k^d}K_{C_j} \right\|  \,\le \, F(m) \,.
\end{equation}
\end{corollary}

\noindent {\em Proof of Lemma \ref{simple}:~}  To simplify notation we set $l=m+2g(m)$ in the proof.  
If  $D =\{ x \in \bZ^d \,; \,  a_i \le x_i < a_i +l\}$ is a cube of side length $l$ and $r\in{\mathbb N}$ such that $r < l/2$ we 
denote  $D_r \,=\,   \{ x \in \bZ^d \,; \,  a_i +r \le x_i < a_i +l-r\}$ the cube of side length 
$l-2r$ centered at the same point as $D$. 

Let us consider two cubes $D \subset D' \subset \bZ^d$. We have 
\begin{eqnarray} 
 \| K_{D'} - K_{D} \| \,&\le& \, 
\sum_{X \subset D' \atop X \not \subset D}  \|\psi_X\| 
\,\le\, \sum_{ x \in D'}  \sum_{X \ni x \atop X \not\subset D} \frac{1}{|X|} 
\|\psi_X\|  \nn  \\ 
\,&\le&\,  \sum_{x \in D' \setminus D_r}  \sum_{X \ni x} \frac{1}{|X|} 
\|\psi_X\|  + \sum_{x \in D_r}   \sum_{X \ni x \atop X \not\subset D} \frac{1}{|X|} 
\|\psi_X\|  \nn \\
\,&\le&\, |D' \setminus D_r| \|\Psi\|  + |D_r| \sum_{X\ni 0 \atop {\rm diam }(X) >r } \frac{1}{|X|}  
\|\psi_X\| \,. \label{bou01}
\end{eqnarray} 
Using \eqref{bou01} we have, for any $r$,  
\begin{eqnarray}
&& \limsup_{n \to \infty} \frac{1}{|\Lambda(n)|} \| K_{\Lambda(n)} - K_{\Lambda( kl )} \|  \nn \\
&& \,\le \, \lim_{n \to \infty} \left[ \frac{|\Lambda(n) \setminus \Lambda_r(kl)|}
{|\Lambda(n)|} \|\Psi\|  + \frac{|\Lambda_r( kl )|}{|\Lambda(n)|} \sum_{X\ni 0 \atop 
{\rm diam }(X) >r } \frac{1}{|X|} \|\psi_X\| \right]  \nn \\
&& \,=\, \sum_{X\ni 0 \atop  {\rm diam }(X) >r } \frac{1}{|X|} \|\psi_X\|  \,. \nn
\end{eqnarray}
Since $r$ is arbitrary we have 
\begin{equation}\label{bbb01}
 \limsup_{n \to \infty} \frac{1}{|\Lambda(n)|} \| K_{\Lambda(n)} - K_{\Lambda( kl )} \| =0 \,. 
\end{equation}
Using \eqref{bou01} again we have 
\begin{eqnarray}
&& \limsup_{n \to \infty} \frac{1}{|\Lambda(n)|} \left\| \sum_{j=1}^{k^d}  \left( K_{{\tilde C}_j} - 
K_{C_j}\right) \right\|  \nn \\
&&\,\le\, \lim_{n \to \infty} \frac{k^d |\Lambda(l)|}{|\Lambda(n)|} \left[ \frac{|\Lambda(l) 
\setminus \Lambda_r(m)|}
{|\Lambda(l)|} \|\Psi\|  + \frac{|\Lambda_r( m)|}{|\Lambda(l)|} \sum_{X\ni 0 \atop 
{\rm diam }(X) >r } \frac{1}{|X|} \|\psi_X\| \right] \,. \nn 
\end{eqnarray}
If  $r= h(m)$ with $\lim_{m \to \infty}h(m)= \infty$ and $\lim_{m \to \infty} h(m)/m \,=\,0$,
we get 
\begin{equation}\label{bbb02}
\limsup_{n \to \infty} \frac{1}{|\Lambda(n)|} \left\| \sum_{j=1}^{k^d}  \left( K_{{\tilde C}_j} - K_{C_j}\right) \right\| \,=\,  o(m) \,.
\end{equation}
Finally 
\begin{eqnarray}
&& \left\| K_{\Lambda(kl)}  - \sum_{j=1}^{k^d} K_{{\tilde C}_j}\right\| \,\le \, 
\sum_{X \subset \Lambda(kl)  \atop X \not\subset {\rm \,\,some\,\, }\tilde{C}_j} \|\psi_X\|  
\,= \,  \sum_{X \subset \Lambda(kl) \atop X \not\subset {\rm \,\,some\,\, }\tilde{C}_j}  
\sum_{j=1}^{k^d}\frac{ | X \cap \tilde{C}_j|}{|X|} \|\psi_X\| \ \nn \\
&& \,  \le  \,  |\Lambda(kl)| \frac{1}{k^d}  \sum_{j=1}^{k^d}  \frac{1}{|\tilde{C}_j|} 
\sum_{X \not \subset \tilde{C}_j}   \frac{|X\cap \tilde{C}_j|}{|X|} \|\psi_X\| 
\,=\, |\Lambda(kl) | d(\Psi, l) \label{bb02}
\end{eqnarray}
with
\begin{eqnarray}
d(\Psi,l)\,&=&\, \frac{1}{|\Lambda(l)|} \sum_{ X \not\subset \Lambda(l)} 
\frac{|X \cap \Lambda(l)|}{|X|} \|\psi_X\| \,=\, \sum_{x \in \Lambda(l)} \sum_{X \ni  x \atop 
X \not\subset \Lambda(l)} \frac{1}{|X| |\Lambda(l)|} \|\psi_X\| \nn \\
 \,&\le&\,  \frac{|\Lambda_r(l)|}{|\Lambda(l)|}  \sum_{X \ni  0 \atop {\rm diam}(X)>r} \frac{1}{|X|} 
\|\psi_X\|   + \frac{|\Lambda(l)| - |\Lambda_r(l)|}{|\Lambda(l)|} \|\Psi\|\,.
\end{eqnarray}
Since $l= m+ 2g(m)$ if we pick $r=h(m)$ as above we get 
\begin{equation}\label{bbb03}
\limsup_{n \to \infty} \frac{1}{|\Lambda(n)|}  \left\| K_{\Lambda(kl)}  - \sum_{j=1}^{k^d} K_{{\tilde C}_j}\right\| \,=\, o(m)
\end{equation}
Combining the bounds \eqref{bbb01}, \eqref{bbb02}, and \eqref{bbb03} concludes the proof 
of Lemma \ref{simple}. \qed

\noindent {\em Proof of Corollary \ref{verysimple}:~}  An easy estimate shows that the difference 
between  $\| |\Lambda(n)|^{-1} K_{\Lambda(n)} - |\Lambda(km)|^{-1} \sum_{j=1}^{k^d}K_{C_j} \|$
and $|\Lambda(n)|^{-1} \| K_{\Lambda(n)} - \sum_{j=1}^{k^d}K_{C_j} \|$  is $O(g(m)/m) \|\Psi\|$.
 \qed

The second fact is a general remark on the strategy to 
prove the existence of a concave RL function \cite{LPS2}

\begin{remark}{\rm \label{rkey}
Let $x, x_1, x_2$  such that $\frac{1}{2}(x_1 + x_2) =x$ and  let  $0 < \varepsilon' < \varepsilon$.   
To prove the existence of a concave RL-function it is enough to prove that 

\begin{equation}\label{mb1}
\underline{m}( B_\varepsilon(x)) \ge \frac{ \overline{m}( B_{\varepsilon'} (x_1)) + 
\underline{m}( B_{\varepsilon'}(x_2))}{2} \,.
\end{equation}
Indeed if we set $x_1=x_2=x$ in \eqref{mb1} then we obtain 
$$
\underline{s}(x) \ge \overline{s}(x) \,,
$$
and therefore the Ruelle-Lanford function $s(x)$ exists.  Using then \eqref{mb1} again we 
obtain that
$$
s(x) \,\ge \, \frac{ s(x_1) + s(x_2)}{2} \,.
$$
Since $s(x)$ is upper-semicontinuous, this implies that  $s(x)$ is concave. 
}
\end{remark}

\subsection{Tracial state and conserved quantities}   In this section we prove a quantum 
large deviation theorem in the simplest possible case.  We bypass a number of  issue 
associated to taking thermodynamic limits for the states  by considering first 
the {\em finite volume  Gibbs states}  
$$
\omega_{\Lambda(n)} ( A) \,=\, \frac{ \tr\left( A e^{- H_{\Lambda(n)}}\right)}{ 
\tr \left(e^{-H_{\Lambda(n)}}\right)}\,.
$$
In addition we assume that the Hamiltonian and that the macroscopic observables $K_\Lambda$ is a 
{\em conserved quantity}, i.e.,  the commutators $[K_\Lambda, H_\Lambda]$ vanish for all $\Lambda$. 
Note that, although very restrictive, this condition is, in general, satisfied for thermodynamic quantities 
such as magnetization, density, energy, etc....  The following theorem provides a (weak)
justification that macroscopic conserved quantities are exponentially concentrated in equilibrium.  

A important special case is the case where $H_\Lambda=0$, that is one consider the tracial state 
${\rm tr}$. In this case any observable $K_\Lambda$ can be chosen arbitrarily and the rate function
$s(x)$ is the microcanonical entropy whose existence is of course  well-known.  The large deviation  
statement for the tracial state can be found e.g. in \cite{Si};  the only novelty here, maybe, is a very simple proof.

\begin{theorem}\label{tracial} Let $\Phi$ and $\Psi$ be interaction with $\|\Phi\| < \infty$ and  $\|\Psi\|< \infty$. 
Suppose that  the commutators $[K_{\Lambda(n)}, H_{\Lambda(n)}]$  commute for all $n$. 
Then the probability measures 
$$
\mu_n (A) \,=\, \frac{ \tr \left(  {\bf I}_A \left( |\Lambda(n)|^{-1} K_{\Lambda(n)}\right) 
e^{-H_\Lambda(n)} \right)}{ \tr \left( e^{-H_\Lambda(n)}\right)} \,,
$$
satisfies a large deviation principle on the scale $|\Lambda(n)|$ with a concave rate function 
$s(x)$. We have 
$$
\sup_{x}( \alpha x + s(x) ) \,=\,  P(\alpha) \,, \quad \quad 
s(x) \,=\, \inf_{\alpha} ( P(\alpha) - \alpha x ) 
$$ 
where 
$
P(\alpha)= \lim_{n\to\infty} |\Lambda(n)|^{-1} \log \tr\left( e^{ - H_{\Lambda(n)} + \alpha K_{\Lambda(n)}}\right) 
$ 
is the translated free energy.
\end{theorem}

\proof 
\noindent Let us choose $x,x_1,x_2$ and $\varepsilon$, $\varepsilon'$ as in Remark \ref{rkey}.   
Given $n >m$  let $k$ be the  even integer such that $n=km +r$ with $0\le r < 2m-1$ 
(haveing $k$ even is useful later).    Divide the  cube $\Lambda(km)$ into  $k^d$ disjoint contiguous cube  
$C_j$, $j=1, \cdots, k^d$ each of which is a translate of  the cube $\Lambda(m)$.   

Let us denote by $\lambda^{(n)}_j$ the eigenvalues of $H_{\Lambda(n)}$ and by $\mu^{(n)}_j$ 
the eigenvalues of $K_{\Lambda(n)}$.   Since $H_{\Lambda(n)}$ and $K_{\Lambda(n)}$ commute 
we have 
\begin{equation}\label{lamu}
\mu_n (B_\varepsilon(x)) \,=\, \frac{ \sum_{ j \,;\, \frac{\mu^{(n)}_j}{|\Lambda(n)|} \in B_\varepsilon(x) } 
e^{- \lambda^{(n)}_j} }{ \sum_{j} e^{-\lambda^{(n)}_j}} \,.
\end{equation}
By Corollary \ref{verysimple} we can choose $M$ and 
$N=N_m$ so that for $m>M$ and $n>N$ we have
$$
\left\||\Lambda(n)|^{-1} K_{\Lambda(n)} - |\Lambda(km)|^{-1} \sum_{j=1}^{k^d} 
K_{C_j} \right \| \le  (\varepsilon - \varepsilon')
$$
Let $\mu^{(m)}$ be an eigenvalue of $K_{\Lambda(m)}$ with $\mu^{(m)}/|\Lambda(m)| \in 
B_{\varepsilon'}(x_1)$ and let ${\hat \mu}^{(m)}$ be an  eigenvalue of $K_{\Lambda(m)}$ with 
${\hat \mu}^{(m)}/|\Lambda(m)| \in B_{\varepsilon'}(x_2)$.  Let us assign $\mu^{(m)}$ to each 
cube $C_j$ with  $j=1, \cdots, \frac{k^d}{2}$ and  ${\hat \mu}^{(m)}$ to the each  cube $C_j$ with  
$j=\frac{k^d}{2} +1, \cdots,  k^d$.  Then ${\tilde \mu}^{(km)} \equiv \frac{k^d}{2}( \mu^{(m)} + 
{\hat \mu}^{(m)})$ is an eigenvalue of $\sum_{j} K_{C_j}$ such that ${\tilde \mu}^{(km)}/|\Lambda(km)| \in 
B_{\varepsilon'}(x)$.  For $m>M$ and $n\ge N=N_m$, by Weyl's perturbation theorem,  for any 
choice of $\mu^{(m)}$ and ${\hat \mu}^{(m)}$ there exists  an eigenvalue $\mu^{(n)}$ of 
$K_\Lambda(n)$  such that $\mu^{(n)}/|\Lambda(n)| \in B_\varepsilon(x)$. 

Assume that the eigenvalues $\lambda_i^{(n)}$ of $H_{\Lambda(n)}$ are listed 
in increasing order, counting multiplicity.  Let ${\tilde \lambda}_i^{(n)}$ be the eigenvalues of 
$\sum_{j} H_{C_j}\otimes 1_{\Lambda(n) \setminus \Lambda(km)}$  also listed in increasing order.
By Weyl's perturbation theorem, and Lemma \ref{simple}, there exists $M'$ such that for
$m> M'$ there exists $ N'= N'_m$ such that $n \ge N'$ we have 
$$
{\tilde \lambda}_i^{(n)} -  |\Lambda(n)| F(m) \,\le \, \lambda_{i}^{(n)} \le   {\tilde \lambda}_i^{(n)} +  
|\Lambda(n)| F(m) \,.
$$ 
Using the formula \eqref{lamu} we obtain that 
\begin{eqnarray}
&&\mu_n( B_\varepsilon(x)) \ge  \mu_m( B_{\varepsilon'}(x_1))^{\frac{k^d}{2}}  \mu_m( B_{\varepsilon'}(x_2))^{\frac{k^d}{2}}  e^{ -2 |\Lambda(n)| F(m) } \nn
\end{eqnarray}
and thus 
\begin{eqnarray}
\frac{\log \mu_n( B_\varepsilon(x))}{|\Lambda(n)|} &\ge&  
\left(  \frac{\log \mu_m( B_{\varepsilon'}(x_1))}{2 |\Lambda(m)|} + 
\frac{\log \mu_m( B_{\varepsilon'}(x_2))}{2 |\Lambda(m)|} \right)  
\frac{k^d |\Lambda(m)|}{|\Lambda(n)|}  -2 F(m) \nn 
\end{eqnarray}
To conclude we take first a $\liminf$ over $n$ keeping $m$ fixed and then choose a subsequence 
$m_l$ such that 
$
\lim_{l\to\infty} |\Lambda(m_l)|^{-1} \log \mu_{m_l}( B_{\varepsilon'}(x_1)) 
= \overline{m}(B_\varepsilon'(x_1)) \,.
$
Together with Remark \ref{rkey} this concludes the proof of Theorem \ref{tracial}. 
\qed

\begin{theorem}\label{tracial+} Let $\Phi$ and $\Psi$ be interaction with 
$\|\Phi\| < \infty$ and  $\|\Psi\|< \infty$.  Suppose that  the commutators 
$[K_{\Lambda(n)}, H_{\Lambda(n)}]$  vanish for all $n$.  Suppose $\omega^{(\Phi)}$ 
satisfies the condition \eqref{decoup0cl}.  Then the probability measure 
$$
\mu_n (A) \,=\, \omega^{(\Phi)} \left(  {\bf I}_A \left( |\Lambda(n)|^{-1} K_{\Lambda(n)}\right)  \right)
$$
satisfies a large deviation principle on the scale $|\Lambda(n)|$ with a concave rate function 
$s(x)$. We have 
$$
\sup_{x}( \alpha x + s(x) ) \,=\,  P(\alpha) \,, \quad \quad 
s(x) \,=\, \inf_{\alpha} ( P(\alpha) - \alpha x ) 
$$ 
where 
$
P(\alpha)= \lim_{n\to\infty} |\Lambda(n)|^{-1} \log \tr\left( e^{ - H_{\Lambda(n)} + \alpha K_{\Lambda(n)}}\right) 
$ 
is the translated free energy.
\end{theorem}

\proof  Since 
$$
\omega^{(\Phi)} \left( {\bf I}_{A}\left(  |\Lambda(n)|^{-1} K_{\Lambda(n)}\right)  \right) \, \ge \, e^{- c(n)} \frac{ \tr \left(  {\bf I}_A \left( |\Lambda(n)|^{-1} K_{\Lambda(n)}\right) e^{-H_\Lambda(n)} \right)}{ \tr \left( e^{-H_\Lambda(n)}\right)} 
$$
the theorem follows immediately from Theorem \ref{tracial}. \qed

\begin{remark}{\bf (Equivalence of ensembles)} {\rm For the tracial case 
it is not difficult \cite{Si} to show the variational formula  $s(x) \,=\, \sup\left\{ s(\omega) \,;\, 
 \omega( A_\Psi) = x \right\}$ where $s(\omega)$ is the specific entropy of the state $\omega$
and that the supremum is attained exactly if $\omega=\omega^{\beta \Phi}$ is a Gibbs-KMS 
state at temperature $\beta=\beta(x)$ with $\beta$ chosen in such a way  that 
$\omega^{\beta \Psi}( A_\Psi)=x$.  This is the equivalence of ensemble: the 
thermodynamic function entropy can be computed via microcanonical or canonical prescriptions. 
Furthermore the LDP can be used to prove that suitable microcanonical states are equivalent 
to canonical states, see \cite{Si} for the classical case and \cite{Li1,Li2} for the quantum case. 
Non-commutative versions of equivalence of ensembles are considered in \cite{DMN}.
}
\end{remark}

\subsection{Classical subalgebras}

In this section we assume that $\omega$ is an asymptotically decoupled state and that 
$\Psi \in \calB$ is a classical interaction, i.e., there exists a classical subalgebra 
${\cal O}^{(cl)} \subset {\cal O}$ such that, for all $X$,  $\psi_X \in {\cal O}^{(cl)}$. 
For example if $\Psi= \{ \psi_x\}_{x \in \bZ^d}$ consists of only of "one-site" interaction 
then $\Psi$ is classical.  More generally any classical spin system is described by a 
classical interaction.  Note that we do not assume any relation  between the interaction 
$\Psi$ and the state $\omega$; if $\omega= \omega^{\Phi}$ is a  Gibbs state for the interaction 
$\Phi$ then $\Phi$ and $\Psi$ need not commute. 

As noted in Section \ref{formalism} the restriction of $\omega$ on ${\cal O}^{(cl)}$ can be 
identified with a probability measure $d\omega^{(cl)}$ on the configuration space $\cal L$. 
Furthermore it is easy to see that the state $\omega^{(cl)}$ on the $C^*$-algebra 
${\cal O}^{(cl)}\simeq C(\calL)$  
is asymptotically decoupled whenever the state $\omega$ on $\cal O$ 
is asymptoticallly decoupled.  




We have 
\begin{theorem}\label{csa} 
Let $\Psi$ be a classical interaction with $\|\Psi\| < \infty$ 
and let $\omega$ be an asymptotically decoupled state. Then the sequence of  
probability measures 
$$
\mu_n (A) \,=\, \omega \left(  {\bf I}_A \left( |\Lambda(n)|^{-1} K_{\Lambda(n)}\right) \right) \,,
$$
satisfies a large deviation principle on the scale $|\Lambda(n)|$ 
with a concave rate function $s(x)$.  Moreover 
$$
s(x) \,=\, \inf_{\alpha}( f(\alpha) - \alpha x)
$$
where 
$$
f(\alpha)\,=\, \lim_{n\to\infty} \frac{1}{|\Lambda(n)|}\log\omega\left( \exp(\alpha K_{\Lambda(n)}) \right) \,.
$$
\end{theorem}

\proof   The proof reduces to the classical case (see \cite{LPS2}) 
since the measures $\mu_n$ can be written as
$$
\mu_n(A) \,=\,     \omega^{(cl)} \left(  {\bf I}_A \left( |\Lambda(n)|^{-1} K_{\Lambda(n)}\right) \right) 
\,=\, \int   {\bf I}_{\{|\Lambda(n)|^{-1} K_{\Lambda(n)} \in A\}}(l) d\omega^{(cl)}(l)  \,.
$$
and the restriction of $\omega^{(cl)}$ on $\calO^{(cl)}$ is 
asymptotically decoupled. Following Remark \ref{rkey} we choose arbitrary   $x, x_1, x_2$ 
such that  $\frac{x_1}{2} + \frac{x_2}{2} =x$ and  $0 < \varepsilon' < \varepsilon$.  We divide 
the cube  $\Lambda(n)$ as explained before Lemma \ref{simple}. 
%
We choose $M$ and $N=N_m$ such that for $m>M$ and $n>N$ 
\begin{equation} \label{es}
 \left\|  
 \frac{ 1} {|\Lambda(n)|}{ K_{\Lambda(n)}} -  \frac{1}{|\Lambda(km)| }
  \sum_{j=1}^{k^d} K_{C_j}  
 \right\|  \le  \varepsilon - \varepsilon'
\end{equation}
Let $l_{C_j}$ be configurations such that  $K_{C_j}(l_{C_j}) /|C_j| \in B_{\varepsilon'}(x_1)$ for  
$1\le j \le \frac{k^d}{2}$ and $K_{C_j}(l_{C_j})/|C_j|  \in B_{\varepsilon'}(x_2)$ for 
$\frac{k^d}{2}+1 \le j \le  k^d$. By  \eqref{es} any configuration  
$l_{\Lambda(n)}$ which coincides with  $l_{C_j}$ on all $C_j$ satisfies  
$K_{\Lambda(n)}(l_{\Lambda(n)})/|\Lambda(n)| \in B_\varepsilon(x)$. 

Therefore using the fact that $\omega^{(cl)}$ is asymptotically decoupled we have the bound
\begin{eqnarray}
&& \omega \left(  {\bf I}_{B_{\varepsilon}(x)} \left(\frac{K_{\Lambda(n)}}{ |\Lambda(n)|}\right) \right) \,=\,  
\int {\bf I}_{\left\{ \frac{K_{\Lambda(n)}}{ |\Lambda(n)|} \in B_\varepsilon(x)\right\}} \, d\omega^{(cl)} \nn \\
&&\ge \int  \prod_{j=1}^{\frac{k^d}{2}} {\bf I}_{\left\{ \frac{K_{C_j}}{ |C_j|} \in B_{\varepsilon'}(x_1)\right\}}   
 \prod_{\frac{k^d}{2}+1}^{k^d} {\bf I}_{\left\{ \frac{K_{C_j}}{ |C_j|} \in B_{\varepsilon'}(x_2)\right\}} 
 d\omega^{(cl)} \nn \\
&& \ge  \left( \int  {\bf I}_{\left\{ \frac{K_{\Lambda(m)}}{ |\Lambda(m)|} \in B_{\varepsilon'}(x_1)\right\}} 
d\omega^{(cl)} \right)^{\frac{k^d}{2}}  
\left( \int  {\bf I}_{\left\{ \frac{K_{\Lambda(m)}}{ |\Lambda(m)|} \in B_{\varepsilon'}(x_2)\right\}} 
d\omega^{(cl)} \right)^{\frac{k^d}{2}}   e^{- c(m)k^d} \nn
\end{eqnarray}

Thus we obtain
\begin{eqnarray}
\frac{\log \mu_n( B_\varepsilon(x))}{|\Lambda(n)|} &\ge&  
\left(  \frac{\log \mu_m( B_{\varepsilon'}(x_1))}{2 |\Lambda(m)|} 
+ \frac{\log \mu_m( B_{\varepsilon'}(x_2))}{2 |\Lambda(m)|} \right)  \frac{k^d |\Lambda(m)|}{|\Lambda(n)|}  \nn \\
&& -  \frac{1}{|\Lambda(n)|} c(m) k^d \,. \nn
\end{eqnarray}
We conclude by taking the $\liminf$ over $n$ and then then choosing a subsequence 
$m_l$ such that 
$
\lim_{l\to\infty}  ( |\Lambda(m_l)|)^{-1} \log \mu_{m_l}( B_{\varepsilon'}(x_1)) = \overline{m}(B_\varepsilon'(x_1)) \,.
$
The identification of the rate function follows from Varadhan's lemma. \qed

\begin{remark}{\rm  One can show (see \cite{Pf,OR} for more details) that the rate function 
satisfies the following variational characterization:   
$$
s(x) \,=\, \sup\{ - h_{cl}(\nu , \omega^{(cl)} ) \,;\, \nu(A_\Psi) = x \}
$$ 
where $h_{cl}$ is the classical relative entropy per unit volume,  
and the  supremum is taken  over all classical translation invariant states. 
}
\end{remark}

\subsection{Dimension 1}

Throughout this section we assume that $d=1$  (so we write $|\Lambda(n)|=n$)
and that $\omega$ is an asymptotically decoupled state,  for example we may assume 
that $\omega$ a KMS-Gibbs state for a finite range interaction.   We  also assume that  $\Psi$ is a {\em finite range} interaction.

The crucial estimate needed to control the effect of non-commutativity is 
an  estimate on the difference between the spectral projections  associated to 
$K_{\Lambda(n)}$ and $\sum_{j=1}^{k} K_{C_j}$ (see section \ref{preliminaries}). 
To prove this we relies on a "cocycle estimate" proved in \cite{Ar1}, which follows 
from the fact that the time-evolution $\tau_t(A)$ of any local observable $A$ 
for a finite-range quantum spin system can be extended to a entire analytic  
function of $t$.  This allows  to  prove the following "exponential version" of Lemma \ref{simple}. 

\begin{proposition}\label{notsosimple}  Let $\Psi$ be a finite range interaction of range $R$ and 
let  $\beta \in \bR$.   Then there exists a function $F_{\beta}(m)= F_{\beta}(m , R, \Psi)$ 
with 
\begin{equation}
\lim_{m \to \infty} F_{\beta}(m) \,=\, 0 \,.
\end{equation}
such that 
\begin{equation} \label{arabounds}
\limsup_{n \to \infty} \frac{1}{n} \log \left\| e^{ \beta K_{\Lambda(n)}} e^{-\beta \sum_{j=1}^{k} K_{C_j} }\right\| 
 \le  |\beta| F_{\beta}(m)   
\end{equation}
\end{proposition}

\proof  The proof is an application of the results in \cite{Ar1},  see in particular Section 4 and 5.   
The basic bound in \cite{Ar1}, section 5, is that 
if $A_X\in \calO_X$ with ${\rm diam}(X)\le R$ then there exists a constant $D(\beta, R, \Psi)$ such that
\begin{equation}\label{basicAR}
\left\|  e^{\beta K_\Lambda(n)}  e^{- \beta (K_\Lambda(n)-A_X)} \right\| \,\le \, 
e^{|\beta| D(\beta, R, \Psi) \|A_X\|} \,.
\end{equation}
The bound \eqref{basicAR} follows from Dyson formula and estimates (uniform in $n$) on 
the dynamics in imaginary time generated by the Hamiltonian  $K_{\Lambda(n)}$.    
To apply these results here we write 
$$ 
K_{\Lambda(n)} = \sum_{j=1}^{k} K_{C_j} + \sum_{X \subset \Lambda(n)  \atop X \not\subset {\rm \,\,some\,\, }{C}_j } \psi_X \,.
$$
Let $t_{X} \in \{0,1\}$ and  let us define the family of 
interpolating Hamiltonians
$$ 
K_{\Lambda(n)}( \{t_X\} )  = \sum_{j=1}^{k} K_{C_j} + \sum_{X \subset \Lambda(n)  \atop X \not\subset {\rm \,\,some\,\, } C_j} t_X \psi_X   \,.
$$
The estimates on the dynamics in \cite{Ar1} are easily seen to be uniform in $\{t_X\}$ 
and so we can apply the bound \eqref{basicAR} iteratively,  changing at each step one $t_X$ 
from $1$ to $0$. Using that $\Psi$ has a finite range $R$  we obtain the bound
\begin{equation*} 
\left\| e^{ \beta K_{\Lambda(n)}} e^{-\beta \sum_{j=1}^{k} K_{C_j} }\right\|  \le  e^{ 
|\beta| D(\beta, R , \Psi) \sum_{X \subset \Lambda(n) \atop  X \not\subset {\rm \,\,some\,\, }{C}_j} \|\psi_X\|   }
\end{equation*}
But the sum over $X$ is now treated exactly as Lemma \ref{simple} and we find 
$F_{\beta}(m)= F(m)D(\beta, R, \Psi)$. \qed

We use this bound to prove an exponential estimates which control how the spectral 
projections change  when we replace $K_{\Lambda(n)}$ by $\sum_{j=1}^k K_{C_j}$.

\begin{proposition}\label{projectionestimate}  
Let $\varepsilon>\varepsilon'>0$. Then for any $\alpha>0$ there exists a function 
$\tilde{ F}_\alpha(m)$ with $\lim_{m\to \infty} \tilde{F}_\alpha(m)=0$ such that 
\begin{eqnarray}
&&\limsup_{n\to\infty} \frac 1n \log 
\left\|
{\bf I}_{B_{\varepsilon'}(x) } \biggl( (mk)^{-1}  \sum_{j=1}^k K_{C_j} \biggr)
{\bf I}_{B_{\varepsilon}(x)^C  }    \biggl( n^{-1} K_{\Lambda(n)} \biggr) 
\right\|  \nonumber \\
&& \le - \alpha\left(  \varepsilon - \varepsilon' - \tilde{F}_{\alpha}(m) \right)
\end{eqnarray}
\end{proposition}

\proof  Let us write
\begin{equation}
K_{\Lambda(n)}=\sum_{i} \mu_i P_i   \,, \quad  \sum_{j=1}^kK_{C_j} \,=\, \sum_{l} \lambda_{l} Q_l \,,
\end{equation}
where $P_i$ and $Q_l$ are rank-one projections and $\mu_i$ and $\lambda_l$ are 
the eigenvalues   of $K_{\Lambda(n)}$ and  $\sum_{j}K_{C_j}$. 
For any $\beta \in \bR$
\begin{eqnarray}
{\bf I}_{B_{\delta}(y)  }   \biggl( n^{-1} K_{\Lambda(n)} \biggr) 
&\,=\,&  \sum_{i ; \frac{\mu_i}{n} \in B_{\delta}(y) } P_i 
\nonumber  \\
&\,=\,&   
 e^{ \beta ( K_{\Lambda(n)}- ny) } \sum_{i ; \frac{\mu_i}{n} \in B_{\delta}(y)}   e^{ -\beta( \mu_i  - ny ) }   P_i 
\nonumber \\
& \,\equiv\, & e^{\beta (K_{\Lambda(n)}- ny) }  V_{\beta, y, \delta} \label{val}
\end{eqnarray}
and 
\begin{eqnarray}
{\bf I}_{B_{\varepsilon'}(x)  }   \biggl( (mk)^{-1} \sum_{j=1}^k K_{C_j} \biggr) 
&\,=\,&  \sum_{l ; \frac{\lambda_l}{mk} \in B_{\varepsilon'}(x) } Q_l 
\nonumber  \\
&\,=\,&   \sum_{l ; \frac{\lambda_l}{mk} \in B_{\varepsilon'}(x)}   e^{ \beta( \lambda_l  - x n) }   Q_l   
e^{ - \beta (\sum_{j}K_{C_j} - xn) }
\nonumber \\
& \,\equiv\, &  W_{\beta, x, \varepsilon'} e^{ - \beta (\sum_{j}K_{C_j} - xn) } \label{wal} \,,
\end{eqnarray}
with the bounds 
\begin{equation}\label{vwbounds}
\|V_{\beta, y, \delta}\| \,\le \, e^{|\beta|  n \delta}   \,, \quad \|W_{\beta, x, \varepsilon'}\| \,\le 
\, e^{|\beta| mk \left( \varepsilon'  + (\frac{n}{mk}-1) |x|\right)} \,.
\end{equation}
If $y> x$ we choose $\beta = \alpha >0$ and using the equation \eqref{val}, \eqref{wal} as well 
as the  bounds \eqref{arabounds} and 
\eqref {vwbounds} we obtain 
\begin{eqnarray}
&&\limsup_{n \to \infty} \frac{1}{n} \log \left\| {\bf I}_{B_{\varepsilon'}(x) } \biggl( (mk)^{-1} 
 \sum_{j=1}^k K_{C_j} \biggr) {\bf I}_{B_{\delta}(y)  }    \biggl( n^{-1} K_{\Lambda(n)} \biggr) 
\right\|  \nonumber \\
&& \,=\,  \limsup_{n \to \infty}\frac{1}{n} \log \left\| W_{\alpha, x, \varepsilon'} 
e^{ - \alpha (\sum_{j}K_{C_j} - nx) }  
e^{\alpha (K_{\Lambda(n)}- ny) }  V_{\alpha, y, \delta} \right\|  \nonumber \\
&& \,\le\, \limsup_{n \to \infty} \left[  - \alpha (y - x) + \frac{1}{n}\log  
\| e^{- \alpha \sum_j K_{C_j} } e^{\alpha K_{\Lambda(n)}} \|     \right.  \nn \\
&& \left. \hspace{3cm} +  \frac{\alpha}{n}  \left( n \delta + mk \varepsilon'  
 + (n-mk) |x|\right) \right] \nonumber \\
&&\,\le \, -\alpha( y - x) + \alpha F_{\alpha}(m) + \alpha(\delta + \varepsilon')  + 
\alpha \frac{g(m)}{m} |x| \nn
\end{eqnarray}
Similarly for $y <x$ we choose $\beta=-\alpha$ and obtain a similar bound and finally 
\begin{eqnarray}
&&\limsup_{n \to \infty} \frac{1}{n} \log \left\| {\bf I}_{B_{\varepsilon'}(x) } \biggl( (mk)^{-1} 
 \sum_{j=1}^k K_{C_j} \biggr)
{\bf I}_{B_{\delta}(y)  }    \biggl( n^{-1} K_{\Lambda(n)} \biggr) 
\right\|  \nonumber \\
&&\,\le\, -\alpha |y - x| + \alpha F_{\alpha}(m)+ \alpha (\delta + \varepsilon') + \alpha \frac{g(m)}{m}|x|
 \label{lbound}
\end{eqnarray}
Next we choose $\delta$ be such that $\varepsilon > 2 \delta + \varepsilon'$ and 
choose finitely many intervals $T_l$ and $x_l \in T_l$, $l=1, \cdots, L$ such that 
$$
B_{\varepsilon}(x)^C \cap \left[ - \| \Psi\|, \|\Psi\| \right]  \,=\,  \cup_{l} T_l \,,\quad T_l \subset B_\delta(x_l)\,.
$$
By the principle of the largest term,  and using the bound \eqref{lbound} we obtain
\begin{eqnarray}
&&\limsup_{n \to \infty}  \frac{1}{n} \log \left\|
{\bf I}_{B_{\varepsilon'}(x) } \biggl( (mk)^{-1}  \sum_{j=1}^k K_{C_j} \biggr)
{\bf I}_{B_{\varepsilon}(x)^C  }    \biggl( n^{-1} K_{\Lambda(n)} \biggr) 
\right\|  \nonumber \\
&& \,\le \, \limsup_{n \to \infty}  \frac{1}{n} \log \left\|
 {\bf I}_{B_{\varepsilon'}(x) } \biggl( (mk)^{-1}  \sum_{j=1}^k K_{C_j} \biggr) \sum_{l=1}^L
{\bf I}_{T_l }    \biggl( n^{-1} K_{\Lambda(n)} \biggr) 
\right\|  \nonumber \\
&& \,\le \, \max_{l}  \limsup_{n \to \infty}  \frac{1}{n} \log \left\|
 {\bf I}_{B_{\varepsilon'}(x) } \biggl( (mk)^{-1}  \sum_{j=1}^k K_{C_j} \biggr) 
{\bf I}_{B_\delta(x_l) }    \biggl( n^{-1} K_{\Lambda(n)} \biggr) 
\right\|  \nonumber \\
&& \,\le\, - \alpha( \varepsilon - \varepsilon' - \delta)  + \alpha \left(F_\alpha(m) +  \frac{g(m)}{m}|x|\right)
\end{eqnarray}
Since $\delta$ is arbitrary this concludes the proof with $\tilde{F}_\alpha(m)= 
F_\alpha(m) + \frac{g(m)}{m}|x|$.    \qed

With this estimate we can now prove

\begin{theorem}\label{1dth} 
Let $d=1$, let $\omega$ be an asymptotically decoupled translation invariant state, and 
let $\Psi$  be a finite range interaction.  
Then the sequence of  probability measures 
$$
\mu_n (A) \,=\, \omega \left(  {\bf I}_A \left( n^{-1} K_{\Lambda(n)}\right) \right) \,,
$$
satisfies a large deviation principle with a concave rate function $s(x)$.  Moreover 
$$
s(x) \,=\, \inf_{\alpha}( f(\alpha) - \alpha x)
$$
where 
$$
f(\alpha)\,=\, \lim_{n\to\infty}  n^{-1} \log\omega\left( \exp(\alpha K_{\Lambda(n)})\right) \,.
$$
\end{theorem}

\proof  Let $\omega$ be an asymptotically decoupled state with parameters $g$ and $c$.Let  $x, x_1, x_2$ be  such that $\frac{x_1}{2} + \frac{x_2}{2} =x$ and  
$0 < \varepsilon'  <  \varepsilon$.  For any $n>m$ we decompose $\Lambda(n)$ 
as in Section \ref{preliminaries}.
Note that 
\begin{equation}\label{prodproj}
\bigotimes_{j=1}^{k/2}   {\bf I}_{B_{\varepsilon'}(x_1)}\left(  \frac{K_{C_j}}{m} \right) 
\bigotimes_{j=k/2 +1}^{k}    {\bf I}_{B_{\varepsilon'}(x_2)} \left(\frac{ K_{C_j}}{m}\right) 
\,\le \, {\bf I}_{B_{\varepsilon'}(x)}\left(  \frac{\sum_{j} K_{C_j}}{mk}\right) \,,
\end{equation}
and that for any projections $P$ and $Q$ and a state $\omega$ we have 
\begin{eqnarray}
\omega(P) \,&=&\, \omega (  Q PQ ) + \omega( (1-Q)P Q  +QP(1-Q) ) + \omega( (1-Q) P (1-Q)) 
\nonumber  \\
\,&\le& \, \omega (Q) + 2\| (1-Q) PQ  \| + \| (1-Q)P(1-Q)\|  \nonumber \\
\,&\le&\, \omega(Q) + 3\|(1-Q) P\| \,. \label{dah}
\end{eqnarray}
Using that $\omega$ is asymptotically decoupled, and  
estimate \eqref{prodproj}--\eqref{dah} we obtain 
\begin{eqnarray}
&&\frac{1}{2m} \log \omega\left(  {\bf I}_{B_{\varepsilon'}(x_1)}
\left( \frac{K_{\Lambda(m)}}{m}\right) \right)
+ \frac{1}{2m} \log \omega\left(  {\bf I}_{B_{\varepsilon'}(x_2)}
\left( \frac{K_{\Lambda(m)}}{m}\right) \right)
\nonumber \\
&& 
\,\le\,  \frac{1}{mk} \log\omega\left( \bigotimes_{j=1}^{k/2}   {\bf I}_{B_{\varepsilon'}(x_1)} 
\left(  \frac{K_{C_j}}{m} \right) \bigotimes_{j=k/2 +1}^{k}   
{\bf I}_{B_{\varepsilon'}(x_2)} \left(\frac{ K_{C_j}}{m}\right) \right) +  \frac{ c(m) k}{mk}
\nonumber \\
&& \,\le\,  \frac{1}{mk} \log \omega \left( {\bf I}_{B_{\varepsilon'}(x)}\left(  \frac{\sum_{j} 
K_{C_j}}{mk}\right)  \right)  +  \frac{ c(m)}{m}
\nonumber \\
&&\,\le\,  \frac{1}{mk} \log   \left[ 
 \omega \left( {\bf I}_{B_{\varepsilon}(x)}\left(  \frac{\ K_{\Lambda(n)}}{n}\right)   \right)  
 \right. 
 \nonumber \\
 && \left.  \hspace{1.8cm}
  + 3 \left\| 
{\bf I}_{B_{\varepsilon'}(x) }     \left( \frac{ \sum_{j=1}^k K_{C_j} }{mk} \right)
{\bf I}_{B_{\varepsilon}(x)^C }  \left(\frac{ K_{\Lambda(n)}}{n}          \right)     
\right\| 
\right] 
 + \frac{ c(m) }{m}   
\nonumber 
 \end{eqnarray}
Keeping $m$ fixed we take a $\liminf$ over $n$ and using  Proposition \ref{projectionestimate}
we obtain
\begin{eqnarray}
&&\frac{1}{2m} \log \omega\left(  {\bf I}_{B_{\varepsilon'}(x_1)}
\left( \frac{K_{\Lambda(m)}}{m}\right) \right)
+ \frac{1}{2m} \log \omega\left(  {\bf I}_{B_{\varepsilon'}(x_2)}\left( \frac{K_{\Lambda(m)}}{m}
\right)\right)
\nonumber \\
&& \le  \left(1 + \frac{g(m)}{m} \right) \max \left\{  {\underline m} (B_\varepsilon(x)) ,  
- \alpha( \varepsilon - \varepsilon' -  \tilde{F}_{\alpha} (m)) \right\} + \frac{c(m)}{m}  
\label{bbo}\,. 
\end{eqnarray}
To conclude we will use the bound \eqref{bbo} repeteadly.  

\vspace{2mm}
\noindent{(a)} Assume first  $x=x_1=x_2$ and assume that $\underline{s}(x) > -\infty$. 
Choose first $\alpha$ so large that $$ 
- \frac{1}{2} \alpha( \varepsilon - \varepsilon')  < \underline{m}(B_\varepsilon(x)) 
$$  
and then $M=M(\alpha)$ so that  $\tilde{F}_{\alpha}(m)\le \frac{1}{2} (\varepsilon - \varepsilon')$ for 
$m>M$.  By \eqref{bbo} we have then
$$
\frac{1}{m} \log \omega\left(  {\bf I}_{B_{\varepsilon'}(x)} \left( \frac{K_{\Lambda(m)}}{m}\right) \right)
\le \left(1 + \frac{g(m)}{m} \right)  \underline{m}(B_{\varepsilon}(x)) + \frac{c(m)}{m}\,.
$$
and thus $\overline{m}(B_{\varepsilon'}(x)) \le \underline{m}(B_{\varepsilon}(x))$.  This implies 
that the Ruelle function $s(x)$ exists and is finite.

\vspace{2mm}
\noindent{(b)} Assume that $\underline{s}(x) > -\infty$ and $x=\frac{1}{2}(x_1+x_2)$.  Repeating the 
same argument as in (a) one obtains, for $m$ large enough, 
\begin{eqnarray}
&&  \frac{1}{2m} \log \omega\left(  {\bf I}_{B_{\varepsilon'}(x_1)}
\left( \frac{K_{\Lambda(m)}}{m}\right) \right)
+ \frac{1}{2m} \log \omega\left(  {\bf I}_{B_{\varepsilon'}(x_2)}\left( \frac{K_{\Lambda(m)}}{m}
\right)\right) \nonumber \\
&& \,\le \, \left(1 + \frac{g(m)}{m} \right)  \underline {m}(B_{\varepsilon}(x)) + \frac{c(m)}{m} \,, \nonumber 
\end{eqnarray}
and this implies that $\frac{1}{2}\overline{m}(B_{\varepsilon'}(x_1)) + 
\frac{1}{2}\underline{m}(B_{\varepsilon'}(x_2)) \le \underline{m}(B_{\varepsilon}(x))$. 
Thus the rate function $s(x)$ is concave wherever it is finite.  

\vspace{2mm}
\noindent{(c)} Let us assume that $\underline{s}(x) = -\infty$.  Then for any $t >0$ 
we can find  $\varepsilon_t$ such that  for $\varepsilon < \varepsilon_t$ we have 
$\underline{m}(B_{\varepsilon}(x)) \le  - t$.  By \eqref{bbo} we have
$$
\frac{1}{m} \log \omega\left(  {\bf I}_{B_{\varepsilon'}(x)} \left( \frac{K_{\Lambda(m)}}{m}\right) \right) 
\le \left(1 + \frac{g(m)}{m} \right) \max\{ -t \,, -\alpha( \varepsilon- \varepsilon' -  \tilde{F}_{\alpha}(m))\}  + \frac{c(m)}{m}
$$
and thus taking $m\to \infty$ we obtain
$$
\overline{m}(B_{\varepsilon'}(x)) \le \max\{ -t\,,\, -\alpha(\varepsilon -\varepsilon') \}
$$
and so
$$
\overline{s}(x) \,\le \, \max\{ -t \,,\, - \alpha \varepsilon \} 
$$
Since $\alpha$ and $t$ are arbitrary we have $\overline{s}(x) =-\infty$. 

\vspace{2mm}
\noindent {(d)} Assume that $\underline{s}(x) = -\infty$ and $x=\frac{1}{2}(x_1+x_2)$.  Repeating the 
same argument as in (c)  for any $t>0$ there exists $\varepsilon_t>0$ such that
for all $\alpha>0$,
\begin{eqnarray}
&&  \frac{1}{2m} \log \omega\left(  {\bf I}_{B_{\varepsilon'}(x_1)}
\left( \frac{K_{\Lambda(m)}}{m}\right) \right)
+ \frac{1}{2m} \log \omega\left(  {\bf I}_{B_{\varepsilon'}(x_2)}\left( \frac{K_{\Lambda(m)}}{m}
\right)\right) \nonumber \\
&& \le 
(1+\frac{g(m)}{m})
\max\{ -t \,, -\alpha
( \varepsilon_t- \varepsilon'-\tilde{F}_{\alpha}(m))\} 
 + \frac{c(m)}{m} \nn
\end{eqnarray}
and this implies that $\frac{1}{2}\overline{m}(B_{\varepsilon'}(x_1)) + 
\frac{1}{2}\underline{m}(B_{\varepsilon'}(x_2)) \le 
\max\{ -t, -\alpha(\varepsilon_t- \varepsilon')\}$. 
Hence we obtain
\[\frac 12 {s}(x_1) +
\frac 12{s}(x_2)=
\frac 12 \overline{s}(x_1) +
\frac 12\underline{s}(x_2)  =-\infty\le
s(x).
\]

Combining (a), (b), (c), and (d) shows the existence of a concave RL-fucntion and this 
concludes the proof of Theorem \ref{1dth}. \qed

\begin{remark}{\rm A characterization of the rate function using classical relative entropies
is proved in \cite{OR}. 
}
\end{remark}


\begin{thebibliography}{99}



\bibitem{Ar1} Araki, H.:
\newblock Gibbs states of a one dimensional quantum lattice.
\newblock Comm. Math. Phys. {\bf 14}, 120--157 (1969)

\bibitem{Ar2} Araki, H.: 
\newblock On the equivalence of the KMS condition and the variational
principle for quantum lattice systems.
\newblock Comm. Math. Phys. {\bf 38}, 1--10 (1974) 

\bibitem{AM} Araki, H.\ and Moriya, H.: 
\newblock Equilibrium statistical mechanics of fermion lattice systems. 
\newblock Rev. Math. Phys. {\bf 15},  93--198 (2003) 




\bibitem{BDKSSR2}  Bjelakovic, I., Deuschel, J.-D.,  Kr\"oger, T., Siegmund-Schultze, R., 
Szkola, A., and Seiler, R.:  
\newblock  Typical support and Sanov large deviations of correlated states. 
\newblock  Comm. Math. Phys.  {\bf 279},  559--584 (2008)


\bibitem{BR} Bratteli, O.\ and Robinson, D.~W.:
\newblock {\em Operator algebras and quantum statistical mechanics}.  Vols.~1 and 2.
\newblock Texts and Monographs in Physics. Berlin: Springer Berlin, 1981


\bibitem{Co} Comets, F.: 
\newblock Grandes d\'eviations pour des champs de Gibbs sur $\bZ^d$. 
\newblock C. R. Acad. Sci. Paris S\'er. I Math. {\bf 303}, 511--513 (1986)


\bibitem{DMN}  De Roeck, W.,  Maes, C., and Neto\v{c}ny, K.: 
\newblock Quantum macrostates, equivalence of ensembles and
an H-theorem. 
\newblock J. Math. Phys. {\bf 47}, 073303 (2006)

\bibitem{DMNR} De Roeck, W.,  Maes, C., Neto\v{c}ny, K., and Rey-Bellet, L.:
\newblock A note on the non-commutative Laplace-Varadhan integral lemma
\newblock  To appear in Rev. Math. Phys.  (2010)


\bibitem{DZ} Dembo, A.\ and Zeitouni, O.: 
\newblock {\em Large deviations techniques and applications}. 
\newblock Second edition. Applications of Mathematics, {\bf 38}. 
New York: Springer, 1998 

\bibitem{DSZ} Deuschel, J.-D., Stroock, D.W., and Zessin, H.: 
\newblock Microcanonical distributions for lattice gases. 
\newblock  Comm. Math. Phys. {\bf 139}, 83--101 (1991)



\bibitem{FNW}  Fannes, M., Nachtergaele, B., and  Werner, R.F.:
\newblock Finitely correlated states on quantum spin chains.
\newblock Comm. Math. Phys. {\bf 144}, 443-490 (1992).


\bibitem{FO} F\"ollmer, H.\ and Orey, S.: 
\newblock Large deviations for the empirical field of a Gibbs measure. 
\newblock Ann. Probab. {\bf 16}, 961--977 (1988)

\bibitem{GLM} Gallavotti, G., Lebowitz, J.~L., and  Mastropietro, V.: 
\newblock Large deviations in rarefied quantum gases.  
\newblock J. Stat. Phys. {\bf 108}, 831--861 (2002)

\bibitem{Ge1} Georgii, H.-O.:
\newblock Large deviations and maximum entropy principle for 
interacting random fields on $ \bZ^d$. 
\newblock Ann. Probab. {\bf 21}, 1845--1875 (1993)

\bibitem{Ge2} Georgii, H.-O.: 
\newblock Large deviations and the equivalence of ensembles for Gibbsian particle systems 
with superstable interaction. 
\newblock Probab Theory Related Fields {\bf 99}, 171--195 (1994)




 
  
 
\bibitem{HMO}
Hiai, F., Mosonyi, M., and Ogawa, T.:
\newblock Large deviations and Chernoff bound for certain correlated states on a spin chain.
\newblock J. Math. Phys {\bf 48},  123301 (2007)


\bibitem{HMOP} Hiai, F., Mosonyi, M,  Ohno, H., and Petz, D.:
\newblock Free energy density for mean field perturbation of states of a one-dimensional 
spin chain.
\newblock Rev. Math. Phys.  {\bf 20}, 335-365 (2008)

\bibitem{HP} Hiai, F. and  Petz, D.:
\newblock  Entropy densities for algebraic states. 
\newblock J. Funct. Anal. {\bf 125}, 287--308 (1994)

\bibitem{Is} Israel, R.~B.: 
\newblock {\em Convexity in the theory of lattice gases.} 
\newblock Princeton series in physics. Princeton: Princeton University
Press, 1979


\bibitem{La} Lanford III, O.~E.: 
\newblock Entropy and equilibrium states in classical statistical mechanics. 
\newblock In {\em Statistical mechanics and mathematical problems}
Lecture Notes in Physics {\bf 20}, Berlin Heidelberg New York:
Springer, 1973, pp.\ 1--113


\bibitem{LLS} Lebowitz, J.~L., Lenci, M., and Spohn, H.: 
\newblock Large deviations for ideal quantum systems. 
\newblock J. Math. Phys. {\bf 41}, 1224--1243 (2000) 


\bibitem{LRB} Lenci, M., and Rey-Bellet, L.:  
\newblock Large deviations in quantum lattice systems: one phase region. 
\newblock J. Stat. Phys. {\bf 119} , 715--746 (2005)






\bibitem{LPS2} Lewis, J.~T., Pfister, C.-E., and Sullivan, W.~G.: 
\newblock Entropy, concentration of probability and conditional limit theorems.
\newblock Markov Process. Related Fields {\bf 1}, 319--386 (1995) 


\bibitem{Li1} Lima, R.:
\newblock Equivalence of ensembles in quantum lattice systems. 
\newblock Ann. Inst. H. Poincar\'e Sect. A {\bf 15}, 61--68 (1971)

\bibitem{Li2} Lima, R.:
\newblock  Equivalence of ensembles in quantum lattice systems: States. 
\newblock Comm. Math. Phys. {\bf 24} 180--192 (1972)

\bibitem{Ma}Matsui,T.:
\newblock  On non-commutative Ruelle transfer operator.
Rev.  Math.  Phys.  {\bf 13}, 1183-1201 (2001)







\bibitem{NR} Neto\v{c}ny, K. and Redig, F.:
\newblock Large deviations for quantum spin systems.
 \newblock J. Stat. Phys. {\bf 117}, 521--547 (2004)




\bibitem{Og}
Ogata, Y.:
\newblock Large deviations in quantum spin chains.
\newblock  Comm. Math. Phys. {\bf 296}, 35--68 (2010)

\bibitem{OR} Ogata, Y. and Rey-Bellet, L.:
\newblock The rate function for quantum large deviaitions. 
\newblock In preparation. 

\bibitem{Ol} Olla, S.:
\newblock Large deviations for Gibbs random fields.
\newblock Probab. Th. Rel. Fields {\bf 77}, 343--357 (1988)


\bibitem{PRV} Petz, D., Raggio, G.~A., and Verbeure A.: 
\newblock Asymptotics of Varadhan-type and the Gibbs variational
principle.
\newblock Comm. Math. Phys. {\bf 121}, 271--282 (1989)


\bibitem{Pf} Pfister, C.-E.:
\newblock Thermodynamical aspects of classical lattice systems. 
\newblock In {\em In and out of equilibrium (Mambucaba, 2000)},
Progr. Probab., {\bf 51}, Birkh\"auser, Boston, 2002. pp.\ 393-472






\bibitem{RZ} Roelly, S. and  Zessin, H.: 
\newblock The equivalence of equilibrium principles in statistical mechanics and some 
applications to  large particle systems. 
\newblock Exposition. Math. {\bf 11},  385--405 (1993)

\bibitem{Ru65} Ruelle, D.: 
\newblock Correlation functionals.  
\newblock J. Math.Phys. {\bf 6}, 201--220 (1965)

 
\bibitem{Ru} Ruelle, D.: 
\newblock {\em Statistical mechanics: rigorous results} 
\newblock River Edge, NJ: World Scientific, 1999 


\bibitem{Si} Simon, B.: 
\newblock {\em The statistical mechanics of lattice gases} Vol.~I.
\newblock Princeton series in physics.
Princeton: Princeton University Press, 1993 



\end{thebibliography}
\end{document}